\documentclass[epj]{svjour}

\usepackage{graphicx}
\usepackage{fancyhdr}
\def\makeheadbox{{%  remove EPJC header
 }}
\def\mytitle{My title} 
\def\myauthors{My name}  
\def\mytype{My type of session}
\def\mysession{My session}
\def\mytitle{Electroweak and Top Physics at the Tevatron and Indirect 
Higgs Limits } %Put your title here!
\def\myauthors{Sandra Leone}%Put your name here!
\def\mytype{Review}
\def\mysession{\myauthors}
\begin{document}
\title{Electroweak and Top Physics at the Tevatron and Indirect Higgs 
Limits}
\author{Sandra Leone\inst{1}  
%\author{Sandra Leone\inst
\thanks{\emph{Email:} sandra.leone@pi.infn.it}%
(for the CDF and D0 Collaborations)
% Second author\inst{2}% etc
%\thanks{\emph{Present address:} INFN Sezione di Pisa (Italy)}%
}                     % Do not remove
\institute{INFN Sezione di Pisa (Italy)}
%\and the second instituteaddress here}
\date{}
\abstract{
We report a selection of the most recent CDF and D0 results 
on top quark and $W$ and $Z$  boson properties, based on Tevatron Run 2 
data. The 
large datasets of $W$ and $Z$ bosons allow  a very precise 
measurement of the $W$ mass and detailed studies of vector boson 
production and asymmetries. Associated production of vector boson pairs 
has been observed and cross sections have been measured. The top quark is 
being studied in great detail, and a 
precision of 1.1\% in the measurement of its mass has been achieved. The 
precise knowledge of top and $W$ masses are constraining the allowed mass 
range of a standard model Higgs in an unprecedented way. 
\PACS{
      {14.70.Fm}{W bosons}   \and  {14.70.Hp}{Z bosons} \and  {14.65.Ha}{Top 
quarks} \and {14.80.Bn}{Standard-model Higgs bosons}
%      {PACS-key}{discribing text of that key}
     } % end of PACS codes
} %end of abstract
\maketitle
\section{Introduction}
\label{intro}
The CDF and D0 experiments are multipurpose detectors taking data at the 
Tevatron Collider. The Tevatron provides proton--antiproton collisions 
at a center-of-mass energy $\sqrt{s}$ = 1.96 TeV. In  2001 the Tevatron 
Run 2 began, 
after a five year period of
significant upgrade of the accelerator itself and of the CDF and D0
experiments. Accelerator performances have kept improving since the start 
of Run 2. A peak luminosity of 2.92 $\times$ 10$^{32}$ 
cm$^{-2}$s$^{-1}$ has been recently achieved, and more than 3 fb$^{-1}$ of 
integrated luminosity has been delivered so far to both experiments. The 
detectors 
collect data with an average efficiency of about  85\%. As of these 
proceedings, $\simeq$ 2.5 
fb$^{-1}$ were written to tape by each experiment. 

A description of the CDF and D0 upgraded detectors can be found in 
~\cite{upgrade}.

\section{$W$ and $Z$ Cross Section Measurements }

$W$ and $Z$ bosons are produced at the Tevatron through $q\bar{q}$
annihilation and are identified by their leptonic decay into 
electrons,
muons and taus. 
The signature is
given by high energy charged leptons and high missing transverse energy
for $W$ candidates and two oppositely charged high energy leptons for $Z$
candidates.  $W$ and $Z$ identification is a key ingredient for top 
physics. $W$ and $Z$  boson decays are often components of background in 
searches for 
processes beyond 
the standard model (SM)  and, being  relatively well known processes, are 
used for calibrations and detector checks. 
The samples of $W$ and $Z$ boson decays collected by CDF and D0 
now number in the  millions of events, and have been used to produce 
excellent 
measurements of electroweak observables.
 
%fin qui
Inclusive cross sections of both $W$ and $Z$ production have been
measured in all the three lepton decay channels~\cite{inclusive}.
%Figure 1 summarizes the CDF and D\O~ cross section measurements. 
All measurements
are in agreement with the NNLO calculations~\cite{nnlo}. The accuracy is 
limited by systematic effects (dominated by the luminosity uncertainty 
of 6\%).
 %Recently a full differential calculation at NNLO became 
%available~\cite{diff_nnlo}. 

\begin{figure}
\includegraphics[width=0.45\textwidth,height=0.35\textwidth,angle=0]{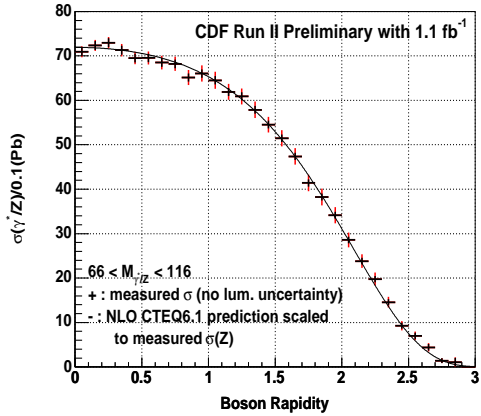}
\caption{The measured $d\sigma/dy$ (crosses) compared to theory 
prediction (solid line) for $Z\rightarrow e^+e^-$.}
\label{fig:1}       % Give a unique label
\end{figure}

The large statistics collected allows CDF 
to produce  a 
$d\sigma(Z)/dy$ measurement for $Z^0/\gamma^* \rightarrow e^+e^-$ events  
obtained from 1.1 
fb$^{-1}$ 
of data. Figure~\ref{fig:1} shows the $d\sigma(Z)/dy$ distribution 
compared to theory prediction. The 
total cross section integrated over all dielectron rapidities is 
$\sigma(Z)$ = 263.34 $\pm$ 0.93(stat)  $\pm$ 3.79(syst) pb. This 
measurement, with 
increased statistics, can be used to constrain the parton distribution 
functions (PDFs).

\begin{figure}
\includegraphics[width=0.45\textwidth,height=0.45\textwidth,angle=0]{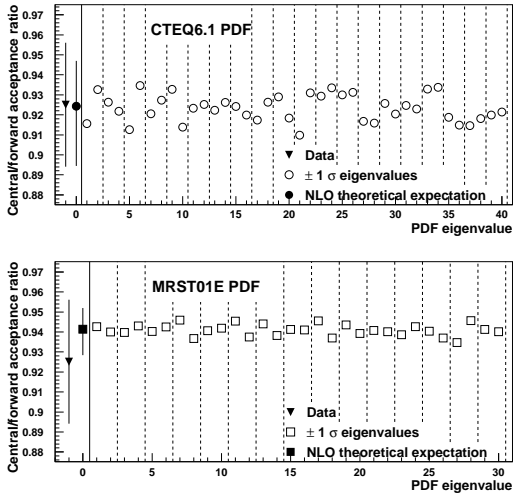}
\caption{CDF experimental ratio of central-to-forward cross sections 
(solid 
triangles) compared
to the CTEQ6.1 (upper plot) and MRST01E (lower plot) acceptance ratios
(solid circles and squares). Dashed lines separate PDF eigenvectors.}
\label{fig:my}
\end{figure}

Recently, CDF measured the 
ratio $R$ of central-to-forward cross sections 
for $p\bar{p} \rightarrow W \rightarrow e\nu$ and obtained $R$ = 0.925 
$\pm$ 0.033~\cite{ratio}. The largest experimental uncertainty, due to 
luminosity, cancels in this ratio. The measurement can be compared to 
theoretical predictions obtained using different PDFs 
(see Fig.~\ref{fig:my}). This quantity is 
sensitive to 
the $W$ rapidity distribution, and provides a novel way to constrain the 
PDFs. 

\section{$W$ Mass Measurement}

The $W$ mass ($M_W$)  is measured in the $e\nu$ and $\mu\nu$ channels from 
a maximum likelihood fit
to the lepton transverse momentum and the transverse mass spectrum, 
defined as: 
$$M_T = \sqrt{2p_T^{\ell}p_T^{\nu} (1 - \cos{\Delta\phi})},$$
where $p_T$ is the lepton transverse momentum and $\Delta\phi$ is the
difference in azimuthal angle between the two leptons.
There
are two main components leading to a precise $M_W$ measurement:
calibration of the detector to the highest possible precision and
simulation of the $p_T$ ($M_T$) spectrum. CDF measured the $W$ mass using 
a sample 
of 200 pb$^{-1}$ of electron and muon data, with the result: $M_W$ = 80413 
$\pm$ 34 (stat) 
$\pm$ 34 (syst) MeV/$c^2$ = 80413 $\pm$ 48 
MeV/$c^2$~\cite{wmass}. This is the most precise single measurement of 
the $W$ mass to date. The updated world average is $M_W$ = 80398 $\pm$ 25 
MeV/$c^2$~\cite{wmass,lep}.
In Table 1  the various contributions to the systematic 
uncertainty are shown. The dominant uncertainties are due to the $W$ boson 
statistics and to the lepton energy scale calibration. They will be 
reduced with 
increased statistics in the $W$ boson and calibration data samples.
% This analysis improves in precision through the
%larger data set  and  also through improved analysis
%techniques and better understanding of the systematic uncertainties. 
Since many simulation parameters are constrained by data control 
samples, their uncertainties are  statistical and are expected to be 
reduced  with more data as well. By the end of Run 2 the Tevatron 
experiments 
should be able to  reduce the uncertainty on $M_W$ below 20 MeV/$c^2$.

\begin{table}
\caption{The uncertainties in MeV/$c^2$ on the $M_T$ fit for $M_W$ 
obtained from 200 pb$^{-1}$ of CDF Run 2 data.}
\label{tab:1}       % Give a unique label
\begin{tabular}{lll}
\hline\noalign{\smallskip}
Uncertainty (MeV/$c^2$)       & Electrons & Muons    \\
\noalign{\smallskip}\hline\noalign{\smallskip}
Lepton scale & 30 & 17 \\
Lepton Resolution & 9 & 3 \\
Recoil Scale & 9 & 9 \\
Recoil Resolution & 7 & 7 \\
$u_{||}$ Efficiency & 3 & 1 \\
Lepton Removal & 8 & 5 \\
Backgrounds & 8 & 9 \\
$p_T$(W) & 3 & 3 \\
PDF & 11 & 11 \\
QED & 11 & 12 \\
Total systematic & 39 & 27 \\
Statistical & 48 & 54 \\
Total & 62 & 60 \\
\noalign{\smallskip}\hline
\end{tabular}
\end{table}

 \section{Direct $W$ Width Measurement}

\begin{figure}
\includegraphics[width=0.45\textwidth,height=0.40\textwidth,angle=0]{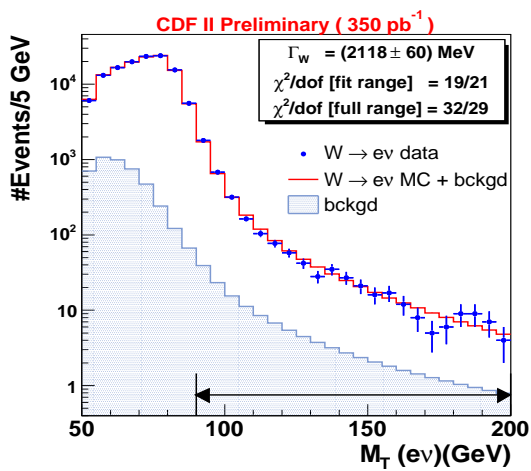}
\caption{The $M_T$ fit for $\Gamma_W$ in $W \rightarrow e\nu$ events. The 
fit is performed in the 
region 90--200 GeV.}
\label{fig:2}       % Give a unique label
\end{figure}

CDF and D0 measured directly the $W$ boson width $\Gamma_W$ using the high 
tail of 
the $M_T$ distribution.
The width is determined by normalizing the predicted signal and background 
$M_T$
distribution in the region of 50 $< M_T <$ 90 GeV/$c^2$ and then fitting
the predicted shape of the candidate events in the tail region 90 $< M_T 
<$
200 GeV/$c^2$ which is most sensitive to the width. CDF has 
the most precise measurement of this quantity, based on 350 pb$^{-1}$ of 
data:
$\Gamma_W$ = 2032 $\pm$ 71 MeV/$c^2$, in good agreement with 
SM  predictions~\cite{wwidth}. 
Figure~\ref{fig:2} shows the $M_T$ distribution in the electron channel 
used for the $\Gamma_W$ measurement. The updated world average is: 
$\Gamma_W$ = 2106 $\pm$ 50 MeV/$c^2$~\cite{wwidth}.

\section{$W$ Charge Asymmetry}

\begin{figure}
\includegraphics[width=0.45\textwidth,height=0.40\textwidth,angle=0]{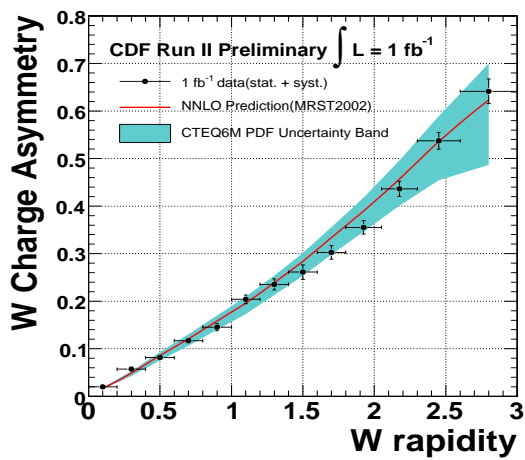}
\caption{Measured $A(|y_W|)$ compared to CTEQ5L prediction. The band 
illustrates the range of uncertainty on the CTEQ prediction.}
\label{fig:wasy}       % Give a unique label
\end{figure}

$W$ bosons at the Tevatron are primarily produced by annihilation of
valence $u$ ($d$) and anti--$d$ (anti--$u$) quarks to $W^+$($W^-$). Since
$u$ quarks carry, on average, a higher fraction of the proton momentum
than $d$ quarks, a $W^+$ ($W^-$) tends to be boosted in the (anti-)proton 
direction. This results in a charge
asymmetry defined as:
$$A_{y_W} = \frac{d\sigma(W^+)/dy_W - d\sigma(W^-)/dy_W}
                 {d\sigma(W^+)/dy_W + d\sigma(W^-)/dy_W},$$
where $y_W$ is the $W$ rapidity and $d\sigma(W^{\pm})/dy_W$ is the
differential cross section for $W^{\pm}$ production. A measurement of the
charge asymmetry is sensitive  to the ratio of $u$ and $d$ quark
components of the PDFs. However, since the
longitudinal component of the neutrino momentum is not measured, the  asymmetry
has been measured traditionally as:
$$A(\eta_e) = \frac{d\sigma(e^+)/d\eta_e - d\sigma(e^-)/d\eta_e}
                   {d\sigma(e^+)/d\eta_e + d\sigma(e^-)/d\eta_e},$$
where $\eta_e$ is the electron pseudorapidity. The observed asymmetry is a
convolution of the $W$ production charge asymmetry and the $V - A$
asymmetry of the $W$ decay~\cite{wasym}. CDF has recently implemented  a 
new analysis method that  
directly reconstructs $y_W$ from $W \rightarrow e\nu$ events, using 1 fb$^{-1}$ 
of 
data. The ambiguity due to the longitudinal neutrino 
component can be partly resolved on a statistical 
basis from the known $ V - A $ decay distribution and 
$d\sigma(W^{\pm})/dy_W$.  
Figure~\ref{fig:wasy} 
shows the measured asymmetry as a function of $y_W$ compared to the CTEQ5L PDF 
prediction.

%input for the next calculation of PDF's.

\section{Diboson Production}

The SM implies that 
the electroweak gauge bosons $W$ and
$Z$ can interact with one another through trilinear and quartic gauge
boson vertices. Study of events containing pairs of vector bosons provides
a sensitive test of the SM since physics beyond the SM could alter the
cross sections and the production kinematics. In addition, diboson 
production represents a test bed for search and  detection of the Higgs 
boson.

All associated production processes involving pairs of $W$, $Z$ and 
$\gamma$ bosons 
have been detected, with cross sections in excellent agreement with SM 
predictions. Both CDF and D0 measured inclusive cross sections for $WW$ 
production ~\cite{ww} and $Z\gamma$ production~\cite{zgamma}. 

\subsection{$W\gamma$ Radiation Amplitude Zero}

\begin{figure}
\includegraphics[width=0.45\textwidth,height=0.35\textwidth,angle=0]{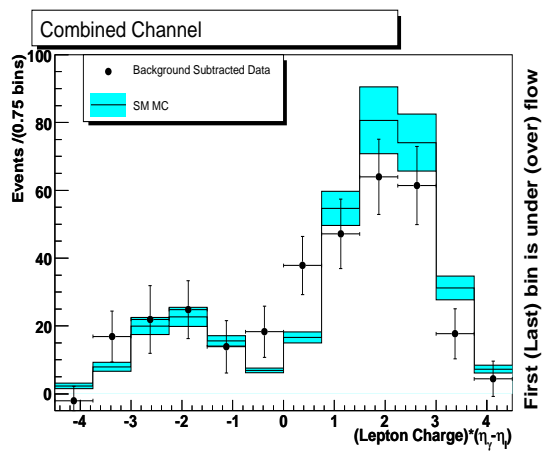}
\caption{D0 result for the charge-signed rapidity difference of $W\gamma$ 
candidates, after  
background subtraction.}
\label{fig:wg}       % Give a unique label
\end{figure}

The $W\gamma$ production can be used to study the gauge structure of the SM. The 
interference among the three tree-level diagrams involved in $W\gamma$ production 
creates a zero in the center-of-mass angular distribution $\theta^*$ between the 
$W$ and the direction of the incoming quarks. D0 measures the charge-signed 
photon-lepton rapidity difference distribution, which is sensitive to the 
radiation amplitude zero. Figure~\ref{fig:wg} shows the charge-signed 
rapidity 
difference in 
both electron and muon channel, obtained by D0 using 900 pb$^{-1}$ of data. The 
observed distribution is consistent with the 
SM prediction and has a shape indicative of the radiation amplitude zero, although 
the result is statistically limited. 

\subsection{$WZ$ Production}

\begin{figure}
\includegraphics[width=0.45\textwidth,height=0.40\textwidth,angle=0]{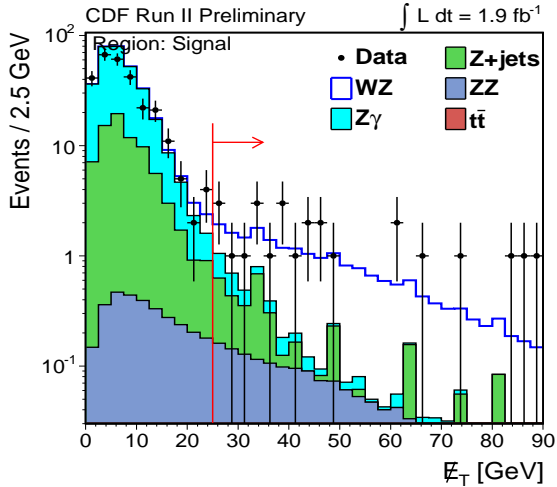}
\caption{Missing $E_T$ distribution of $WZ$ candidates compared to MC 
expectations. 
The arrow indicates the signal region.}
\label{fig:wz}       % Give a unique label
\end{figure}

At $\sqrt{s}$ = {1.96} TeV, the SM predicts $\sigma(WZ)$= 3.7 $\pm$ 
0.25 pb~\cite{theo_wz}.
D0 recently updated on 1 fb$^{-1}$ of data its previous evidence for 
$WZ$ production in events with three charged leptons~\cite{wz_d0}. They 
measure $\sigma (WZ)$ = $2.7^{+1.7}_{-1.3}$ pb. 
% In summer 
%2006 D0 presented the first evidence for $WZ$ production, based on 
%the observation of 12 events in 800 pb$^{-1}$ of data, with a 
%significance of 3.3$\sigma$ and measured $\sigma (p\bar{p} \rightarrow 
%WZ)$ = 3.98 $^{+1.91}_{-1.53}$ (stat+syst) pb. Figure xx shows the transverse 
%mass distribution of candidate events. 
In winter 2007 CDF presented the 
observation of the $WZ$ process based on 1.1 fb$^{-1}$ of 
data~\cite{wz_obs}. In this analysis  
a significant improvement was obtained by exploiting all the available 
detector information in defining  leptons, therefore increasing the lepton  
acceptance. 
Recently CDF measurement was updated on 1.9 fb$^{-1}$ of data.
The measured cross section is $\sigma (WZ)$ =  4.3 
$^{+1.4}_{-1.1}$ pb. 
Figure~\ref{fig:wz} shows the missing $E_T$ distribution in the signal 
region.

%%\begin{figure}
%\includegraphics[width=0.45\textwidth,height=0.40\textwidth,angle=0]{wz_trans_mass.eps}
%\caption{Transverse mass distribution of $WZ$ candidate events.}
%\label{fig:4}
%\end{figure}

\subsection{$ZZ$ Production}

\begin{figure}
\includegraphics[width=0.45\textwidth,height=0.40\textwidth,angle=0]{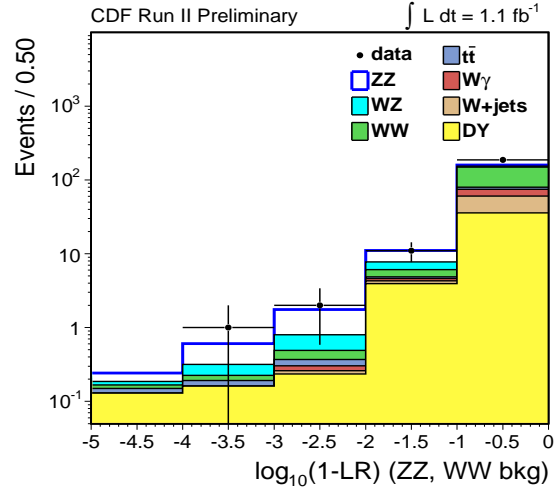}
\caption{Distribution of log(1-$LR$) for $ll\nu\nu$ candidate events used in  
$ZZ$ cross 
section measurement.}
\label{fig:zz}
\end{figure}

The $ZZ$ production cross section predicted by the SM at the Tevatron is
$\sigma(ZZ)$ = 1.4$\pm$0.1 pb at NLO. D0 observed 1 candidate event and 
put an upper 
limit on the production cross section of $\sigma(ZZ)$ $<$ 4.3 pb at 95\% 
C.L..
CDF combined the final states with 4 charged leptons and 2 charged leptons 
plus 2 neutrinos, and did a measurement of the $ZZ$ production 
cross section  $\sigma(ZZ)$ = 0.75 $^{+0.71}_{-0.54}$, based on 1.5 
fb$^{-1}$ of 
data. The observed signal has 
a significance of 3$\sigma$. This is the smallest cross 
section measured at the Tevatron. Figure~\ref{fig:zz} shows the likelihood 
ratio 
distribution for $l l \nu\nu$ candidate events.

\section{Top Quark Physics}

The top quark, discovered in 1995 at the Tevatron~\cite{top_disc}, has 
proven to be a very
interesting particle. It is unique among known fermions because of its 
large mass, of the order of the electroweak symmetry breaking scale. Its 
properties allow to perform stringent 
tests of the SM and to search for new physics through a deviation from SM
predictions. 

At the Tevatron center of mass energy  top quarks are
 produced primarily in $t\bar{t}$ pairs via the strong process $p\bar{p} 
\rightarrow 
t\bar{t}$.
In the SM each top quark decays through charged current weak 
interaction almost exclusively into a
real
$W$ and a $b$ quark ($t \rightarrow Wb$).  Each $W$ subsequently decays
into
 either a charged lepton and a neutrino or two quarks. The
$t\bar{t} \rightarrow W^{+} b W^{-} \bar{b}$ events can thus be identified
by means of different combinations of energetic leptons and jets.
The branching ratio for both $W$'s from a $t\bar t$ pair to decay
leptonically is: 2/81 for $e\mu$, $e\tau$, $\mu \tau$ and  1/81 for $ee$,
$\mu \mu$, $\tau \tau$ ({\it dilepton channels}).
Decay modes of $t\bar t$ pairs in which one $W$ boson decays hadronically
and the other leptonically into an $e$ or a $\mu$ ({\it single lepton + 
jets channel}) have a
branching ratio of 24/81.  When both $W$'s decay hadronically ({\it all
hadronic channel}) the branching ratio is 36/81.
CDF and D0 identified  top quark candidate events using most of these 
signatures.

\section{Top Pair Cross Section Measurement}

\begin{figure*}
\includegraphics[width=0.5\textwidth,height=0.55\textwidth,angle=0]{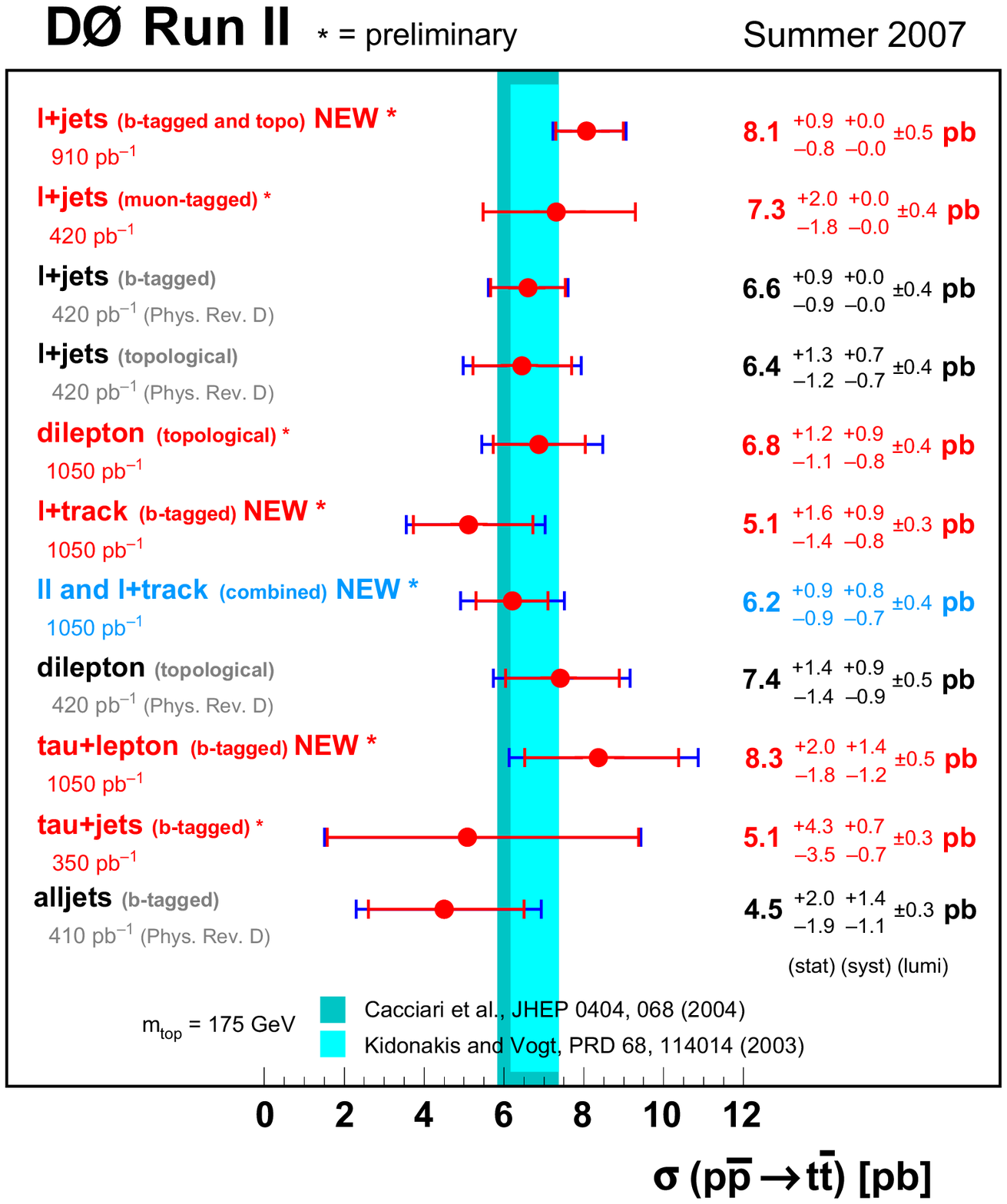}
\includegraphics[width=0.5\textwidth,height=0.55\textwidth,angle=0]{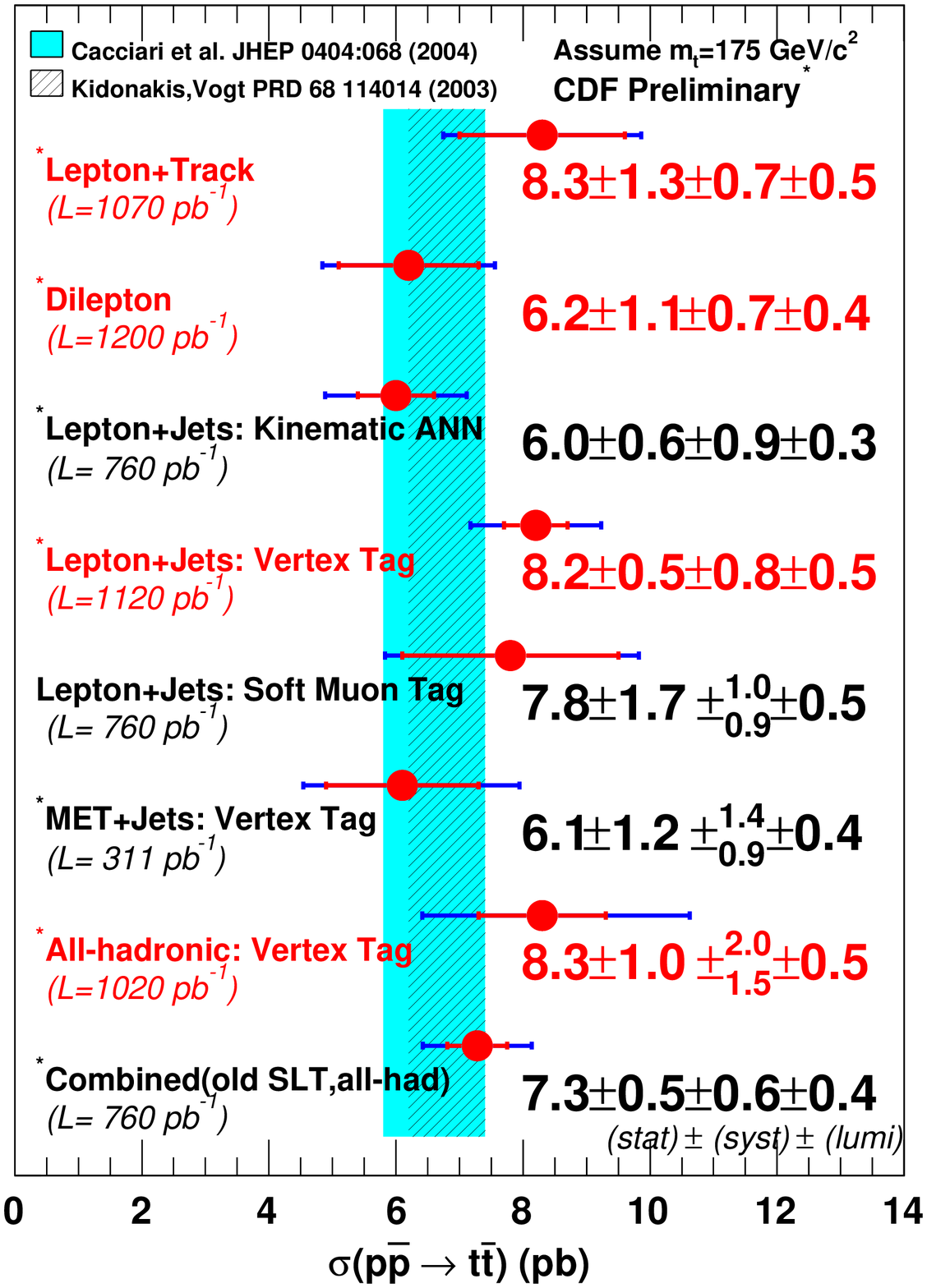}
\caption{A compilation of $t\bar{t}$ production cross section measurements 
at 
$\sqrt{s}$ = 1.96 TeV for $M_{top}$ = 175 GeV/$c^2$, performed by the D0 
(left) and CDF (right) collaborations.}
\label{fig:topxs}
\end{figure*}

By measuring the  $t\bar{t}$ production cross section
$\sigma_{t\bar{t}}$ in many channels and comparing it to perturbative
QCD calculations, we can test the SM
predictions in great detail. The experimental uncertainty on the top quark 
pair production cross section has become comparable to the 
 theoretical one ($\approx$ 12 
\%)~\cite{top_xs}. 
Figure~\ref{fig:topxs}
shows a summary of the top pair cross section measurements in the 
various channels at D0 
(left plot) and CDF (right plot). All the measurements are consistent with 
each other and with the theoretical expectations, which are indicated by 
the vertical band.

\begin{figure}
\includegraphics[width=0.45\textwidth,height=0.40\textwidth,angle=0]{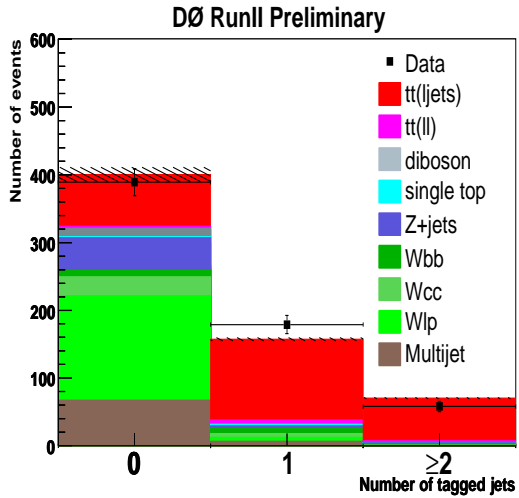}
\caption{Predicted and observed number of events in the zero, single and 
double $b$-tagged single lepton + jets samples, for top candidate events with at 
least four jets.}
\label{fig:012}
\end{figure}

\begin{figure}
\includegraphics[width=0.45\textwidth,height=0.40\textwidth,angle=0]{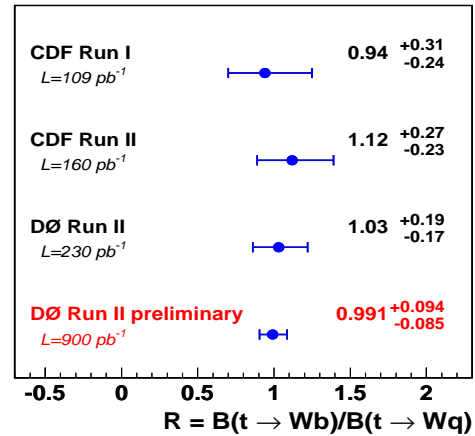}
\caption{Summary of $R$ = $B(t\rightarrow
Wb)/B(t\rightarrow Wq)$ measurements at the Tevatron}
\label{fig:r}
\end{figure}

D0 recently performed a simultaneous measurement of the $t\bar{t}$ 
production cross section and ratio $R$ = $B(t\rightarrow 
Wb)/B(t\rightarrow Wq)$, counting the number of events with 0, 1 and at 
least 2 reconstructed $b$-quark jets (shown in figure~\ref{fig:012}). 
Figure~\ref{fig:r} shows a summary 
of $R$ 
measurements at the Tevatron. The measured $R$ = 
0.991$^{+0.094}_{-0.085}$ (stat+syst) can be translated into a lower limit 
on the  $V_{tb}$ Cabibbo-Kobayashi-Maskawa (CKM) matrix element of 
$|V_{tb}| 
> 0.901$ at 95\% C.L. and assuming CKM unitarity.

\section{Top Mass Measurement}

The top quark mass $M_{top}$ is a fundamental parameter of the SM.
Precise measurements of the top quark and $W$ boson masses constraint the
mass of the Higgs boson. 

The reconstruction of the top quark mass presents several experimental 
challenges.
 The 
neutrinos from leptonically decaying $W$'s escape the detector. The quarks 
hadronize and form jets of particles whose energy must be corrected back to the 
parton level (the precision of the jet energy scale is crucial in this respect). 
The assignment of jets to partons usually has many possible 
permutations. Finally, there are background processes which mimic $t\bar{t}$ 
events. 

\begin{figure}
\includegraphics[width=0.45\textwidth,height=0.40\textwidth,angle=0]{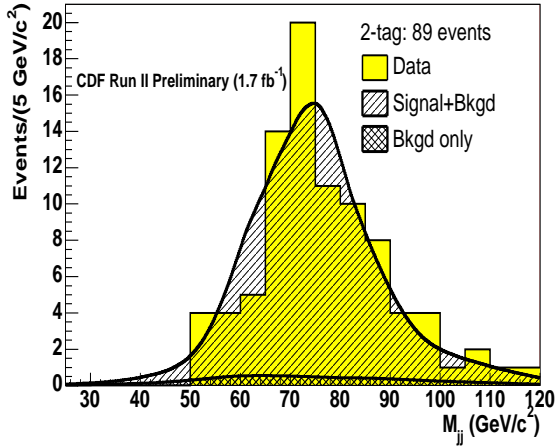}
\caption{ Distribution of the reconstructed dijet mass from single lepton 
+ jets top candidate events with 
two $b$ tagged jets. A fit to signal and background 
templates is overlaid.} 
\label{fig:wjj}
\end{figure}

CDF and D0 performed  many determinations of $M_{top}$, using different 
techniques and all 
the top decay final states. 
In the single lepton + jets and all hadronic channels the uncertainty from 
jet energy scale 
(JES) can be reduced by using 
the reconstructed invariant dijet mass of the 
hadronically decaying $W$ boson in top candidate events as an internal 
constraint (see Figure~\ref{fig:wjj}). This method converts the 
dominant systematic uncertainty into a statistical uncertainty, which will 
improve with more data. 

At the time of this writing, CDF obtained  the most precise determination 
of the top mass in the 
single lepton + jets channel, using a matrix element integration method 
for 
the signal and a neural network discriminant to identify background 
events. CDF finds $M_{top}$ = 172.7 $\pm$ 1.3 (stat.) $\pm$  1.2 (JES) $\pm$ 
1.2 (syst) 
GeV/$c^2$ = 172.7 $\pm$ 2.1 (total) GeV/$c^2$, using 1.7 fb$^{-1}$ of 
data.
The precision of this single measurement is already better than the last 
combined CDF top mass result obtained using up to 1 fb$^{-1}$ of data: 
$M_{top}$ = 170.5 $\pm$ 2.2 (total) GeV/$c^2$.

D0 obtains its most precise top quark mass measurement by combining measurements
performed in the dilepton, single lepton + jets and all hadronic channels. D0
finds: $M_{top}$ = 172.1 $\pm$ 1.5 (stat) $\pm$ 1.9 (syst) GeV/$c^2$ =
172.1 $\pm$ 2.4 (total) GeV/$c^2$, based on up to 1 fb$^{-1}$ of
data.

CDF obtained the best top mass measurement in the dilepton channel using 
the matrix element method and analyzing 1.8 fb$^{-1}$ of data: 
$M_{top}$ = 170.4 $\pm$ 3.1 (stat.) $\pm$ 3.0 (syst) GeV/$c^2$.

\begin{figure}
\includegraphics[width=0.45\textwidth,height=0.40\textwidth,angle=0]{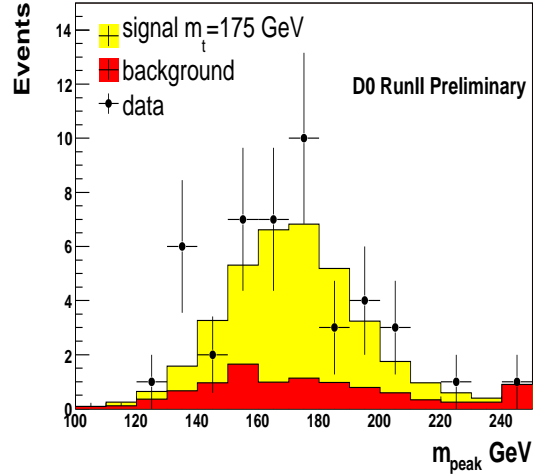}
\caption{Comparison of peak masses in data and Monte Carlo dilepton top 
candidate events.}
\label{fig:dilepton}
\end{figure}

D0 recently presented a new top mass measurement in the dilepton channel based
on 1 fb$^{-1}$ of data,  using two different weighting methods:
$M_{top}$ = 173.7 $\pm$ 5.4 (stat.) $\pm$ 3.4 (syst) GeV/$c^2$.
Figure~\ref{fig:dilepton} shows  a comparison between data and Monte Carlo of
the peak mass for the 57 D0 dilepton top candidate events found in 1 fb$^{-1}$
of data.

\begin{figure}
\includegraphics[width=0.45\textwidth,height=0.60\textwidth,angle=0]{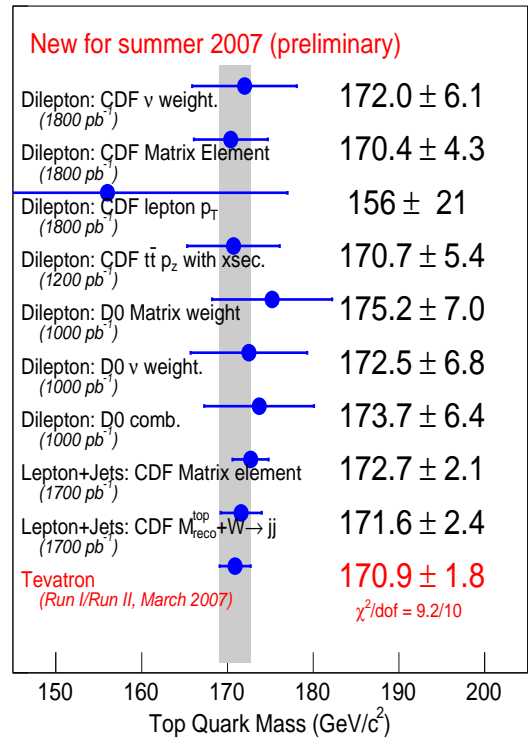}
\caption{A compilation of the most recent CDF and D0 top quark mass 
measurements and last Tevatron combined result.}
\label{fig:topmass}
\end{figure}

 Figure~\ref{fig:topmass} summarizes the most recent CDF and 
D0 top mass measurements and the Tevatron combined top mass 
result: $M_{top}$ = 170.9 $\pm$ 1.1(stat) $\pm$1.5(syst) GeV/$c^2$ = 
170.9 $\pm$ 1.8 (total) GeV/$c^2$, based 
on data-sets including up to 1 fb$^{-1}$ of data~\cite{topmass}. The 
top quark  mass is known with a precision that was thought to be 
unreachable at the 
Tevatron only a few years ago: $\Delta M_{top}/M_{top} \approx 1.1$\%, of 
the order of the top natural width. Therefore, both experiments are now 
addressing a number of effects that, too small to have an impact on the 
previous measurements, could now become important. At the same time, they 
are figuring out which theoretical aspects are relevant, at the 1 
GeV/$c^2$ 
level, and whether they are sufficiently well under control. Before the 
end of Run 2 the Tevatron  experiments are likely to reach a 1 GeV/$c^2$ 
precision on the top quark mass. 

\section{Top Quark Properties}

After the top discovery phase, CDF and D0  moved to detailed studies
of its properties. Both experiments investigated the top candidate events
kinematic properties and the decay vertex. Among the many performed 
studies, here we show a very recent result on the $W$ helicity in top 
decays. More results on top quark properties can be found in~\cite{cabrera}.

\subsection{$W$ Helicity in Top Decays}

\begin{figure}
\includegraphics[width=0.45\textwidth,height=0.40\textwidth,angle=0]{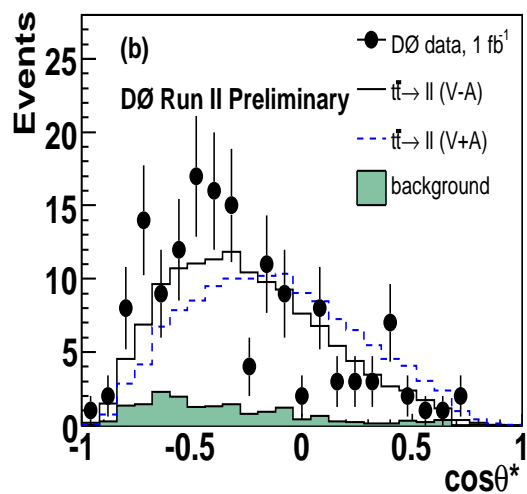}
\caption{$\cos{\theta^*}$ distribution observed in dilepton candidate events at 
D0. The SM prediction is shown as the solid line.}
\label{fig:held0}
\end{figure}

\begin{figure}
\includegraphics[width=0.45\textwidth,height=0.40\textwidth,angle=0]{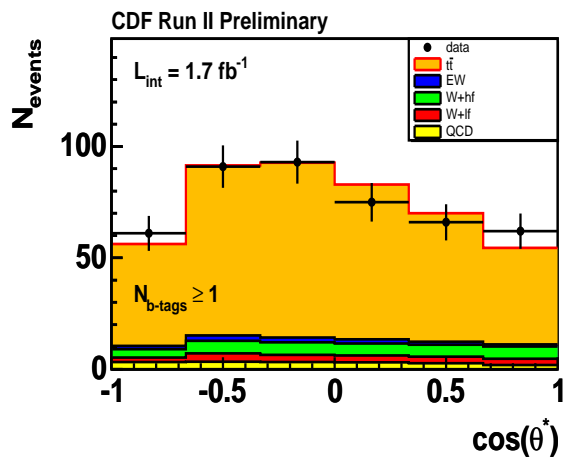}
\caption{The reconstructed $\cos{\theta^*}$ distribution observed in single 
lepton candidate 
events at CDF, together with the SM expectations.} 
\label{fig:helcdf}
\end{figure}

$W$ helicity in top decays is fixed by the $V - A $ structure of the $tWb$ 
vertex and it is reflected in the kinematics of $W$ decay products. SM 
predicts that the fraction of left-handed $W$s is  $F_-$ $\approx$ 30\%, 
the fraction of longitudinally polarized $W$s is $F_0$ $\approx$ 70\%, 
while the right handed fraction $F_+$ is suppressed. Both experiments 
measures the angular distribution of charged leptons in 
the $W$ rest frame measured with respect to the direction of motion of the $W$ 
boson 
in the top-quark rest-frame ($\cos{\theta^*}$). 
Figure~\ref{fig:held0} and~\ref{fig:helcdf} show the $\cos{\theta^*}$ 
distributions observed in D0 dilepton and CDF single lepton + jets 
candidate events respectively.

Using 1 fb$^{-1}$ of single lepton + jets and 
dilepton candidate events D0 measures $F_+$ = 0.017 
$\pm$ 0.048(stat) $\pm$ 0.047(syst) ($F_+$ $<$ 0.14 at 95\% C.L.), with 
$F_0$ fixed to the SM value.

Using 1.7 fb$^{-1}$ of single lepton + jets candidate events CDF measures $F_+$ = 
0.01 $\pm$ 0.05 (stat) $\pm$ 
0.03 (syst) ($F_+$ $<$ 0.12 at 95\% C.L.). CDF attempted also a 2 parameters 
fit, obtaining simultaneously both fractions: $F_0$ = 0.38$\pm$ 0.22 (stat) 
$\pm$ 0.07 (syst) 
and
$F_+$ = 0.15 $\pm$0.10 (stat) $\pm$0.04 (syst). Recently CDF presented a 
new analysis which measures $F_+$ = $-0.04$ $\pm$ 0.04 (stat) $\pm$ 0.03 
and 
therefore extracts a more stringent upper limit: $F_+$ $<$ 0.07  at 95\% 
C.L.. All results are consistent with SM predictions within the 
uncertainties.

\section{Evidence for Single Top Production}

The single top production mechanism involves electroweak production of a
 top quark via the $Wtb$ vertex ($t$ and $s$ channel exchange of a virtual $W$ 
boson). The experimental signature consists of the $W$ 
decay products plus two or three jets, including one $b$ quark jet from the 
decay of the top quark. In $s$-channel events a second $b$ quark jet comes from 
the $Wtb$ vertex. In $t$-channel events a second jet originates from the 
recoiling light-quark and a third low-$E_T$  jet is produced at larger 
$\eta$  
through the splitting of the 
initial state gluon into a $b\bar{b}$ pair. 

The production cross section is predicted 
to be 0.88 and 1.98 pb in the $s$ and $t$ channels respectively~\cite{single} 
for $M_{top}$ = 175 GeV/$c^2$,  
about  half  
than the pair production and with a much larger background. 
On the other 
hand, 
this mechanism allows a direct access to the $V_{tb}$ CKM matrix element, 
and can be used to test the $V - A$ structure of the top charged current 
interaction. 

\begin{figure}
\includegraphics[width=0.45\textwidth,height=0.45\textwidth,angle=0]{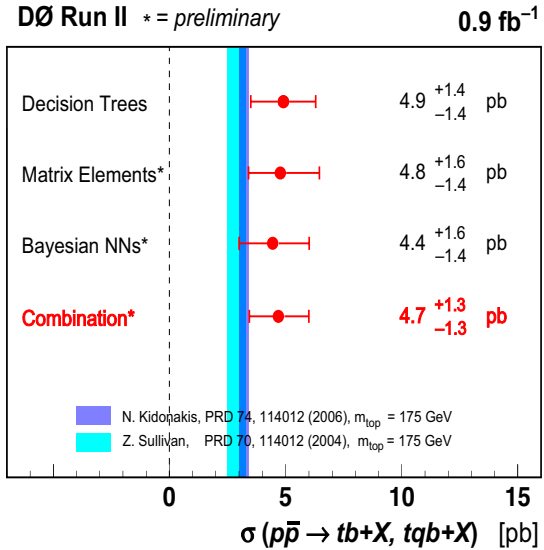}
\caption{D0 evidence for single top production.}
\label{fig:single}
\end{figure}

In order to extract the single top signal from the challenging background 
dominated dataset, both experiments use various multi variate techniques. 
D0 presented the 
first evidence of single top quark production using 0.9 fb$^{-1}$ of 
data~\cite{d0single}. 
D0 searched for single top with three analysis methods: decision trees (DT), 
matrix 
element (ME) and a neural network (NN). Discriminants are constructed with 
a large number of kinematic observables (DT, NN) or by evaluating the 
differential probability of signal using the single top ME. Combining the 
three analyses D0 finds a 3.6$\sigma$ signal and measures a production 
cross section of 4.7 $\pm$ 1.3 pb. This can be translated into the first 
direct measurement of the $V_{tb}$ CKM matrix element: $V_{tb}$ = 1.3 
$\pm$ 0.2 (or  $V_{tb}$: 0.68 $< |V_{tb}| <$ 1 at 95\% C.L.).
Figure~\ref{fig:single} shows the D0 results obtained with the three 
methods.

Recently CDF confirmed the evidence for single top production in 1.5 
fb$^{-1}$ of 
data,  using a multivariate likelihood function technique (giving a 
2.7$\sigma$ excess over the SM background) and a matrix element 
discriminant technique (giving a 3.1$\sigma$ excess). CDF measures: 
$V_{tb}$ = 1.02 $\pm$ 0.18 (exp) $\pm$ 0.07 (theory).

\section{Indirect Limits on Higgs Mass}

\begin{figure}
\includegraphics[width=0.45\textwidth,height=0.45\textwidth,angle=0]{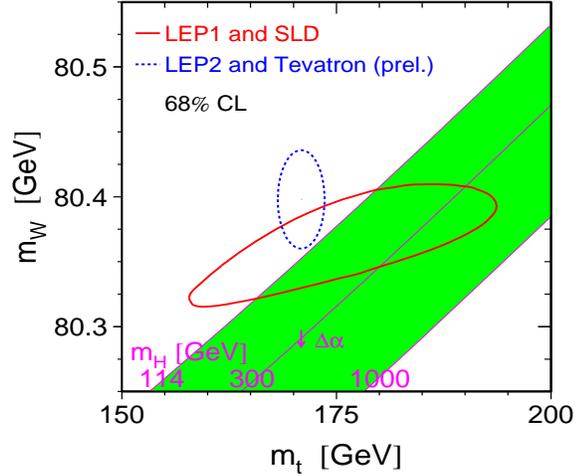}
\caption{Expectations for $M_W$ as a function of $M_{top}$. The green diagonal band 
covers a wide range of Higgs-boson masses. The small ellipse represents 
the combination of LEP2 and Tevatron direct measurements.}
\label{fig:lep1}
\end{figure}

\begin{figure}
\includegraphics[width=0.45\textwidth,height=0.45\textwidth,angle=0]{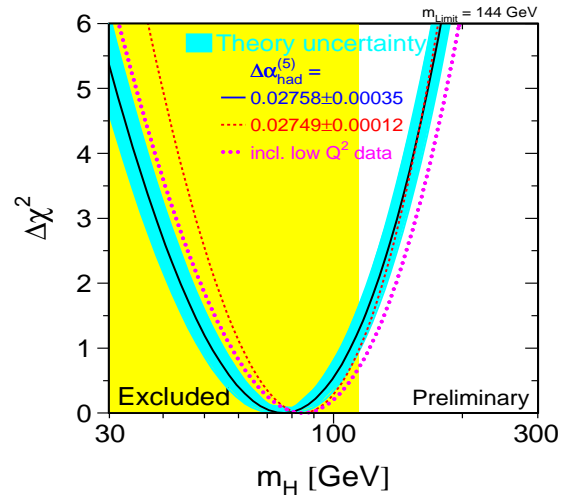}
\caption{$\Delta\chi^2$ curve  as a function of the Higgs-boson mass. The 
minimum of the curve gives the preferred Higgs-boson mass.} 
\label{fig:lep2}
\end{figure}

The Higgs boson is the last remaining SM particle to be observed, and the 
one responsible for generating the $W$ and $Z$ boson masses. Direct 
searches at LEP experiments have excluded a Higgs boson with mass less 
than 114.4 GeV/$c^2$ at 95\% C.L. in the production mode $e^+e^- 
\rightarrow ZH$~\cite{lephiggs}.
The mass of the $W$ depends on the top quark  and Higgs masses through 
radiative effects. Since the Higgs mass is unknown, experimental  
measurements of the top and 
$W$ boson masses provide the strongest indirect constraints on  the Higgs mass, 
based on its contribution to the 
radiative correction  which grows logarithmically with the Higgs mass at 
the one loop level. 
%Global fits to electroweak observables in the SM framework point to a 
%very 
%small value of the Higgs boson mass. 
Figure~\ref{fig:lep1} shows the SM 
prediction of $M_W$ as a function of $M_{top}$ for Higgs masses ranging 
from 
114 
to 1000 GeV/$c^2$. 
%relation of 
%the Higgs mass prediction to the $W$ and top mass measurements using 
%the precision 
%electroweak data. 
%This trend 
%is largely due to the precise measurements of the top quark and $W$ boson masses (see 
%figure~\ref{fig:lep1}). 
Figure~\ref{fig:lep2} shows 
the $\Delta\chi^2$ curve derived from a global fit to precision 
electroweak measurements 
as a function of the Higgs-boson mass, assuming the SM to 
be the correct theory of nature. The preferred value for the Higgs mass, 
corresponding to the minimum of the curve, is  $M_H$ = $76 ^{+33}_{-24}$ 
GeV/$c^2$, well below 
the lower experimental limit set by the LEP 2 experiments.
%, of $M_H$ $>$ 
%114 GeV/$c^2$~\cite{lep}. 
The upper limit on $M_H$ has been set at 144 GeV/$c^2$, at 95\% C.L., and 
 rises to 182 GeV/$c^2$ if one takes into account the LEP 2 direct 
limit.

\begin{figure}
\includegraphics[width=0.45\textwidth,height=0.38\textwidth,angle=0]{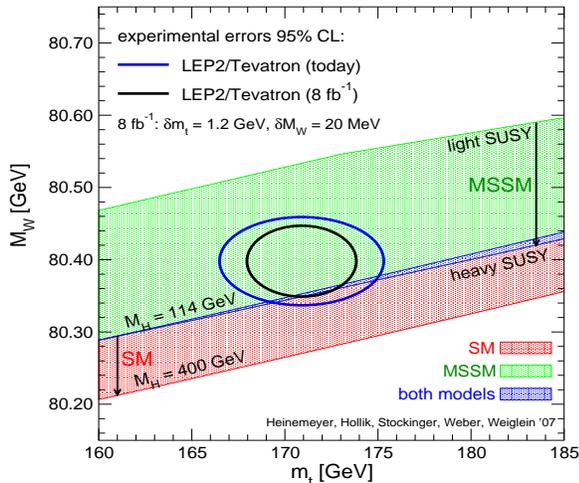}
\caption{Prediction for $M_W$ in the SM and the MSSM as a function of 
$M_{top}$. The allowed MSSM band is obtained scanning over the SUSY mass 
parameters. The allowed SM band is obtained varying the $M_H$ in SM. The 
two 
ellipses correspond to the present experimental situation at 95\% C.L. and 
to the extrapolation assuming 8 fb$^{-1}$ of data at the Tevatron. } 
\label{fig:sven} 
\end{figure}

In the context of the minimal supersymmetric models (MSSM) the 
overall 
agreement 
of all observables appears very good for a wide region of the parameter 
space~\cite{sven}. Figure~\ref{fig:sven} shows the $M_W - M_{top}$ plane  
prediction 
of the SM and the 
MSSM compared to the experimental result. The predictions within the two 
models consist of two bands with a small overlap region. The latter 
corresponds in the SM to a light Higgs boson and in the MSSM to the 
parameters region where all superpartners are heavy. The current 
experimental measurements of $M_W$ and  $M_{top}$ prefer a relatively 
light Higgs mass. Only  by the end of Run 2 one might gather indirect 
information on the Higgs 
SUSY sector from the $M_W$ and $M_{top}$ measurements~\cite{sven}.

\section{Conclusions}

The Run 2 of the Tevatron is well underway. Tevatron experiments have 
in their hands a gold mine of more than 2 fb$^{-1}$ of data.  
Both CDF and 
D0 are producing interesting results in the electroweak 
sector, bringing SM tests to a level of precision  which meets or 
exceed that of electron-positron colliders. The top quark mass is 
known with a 1.1\% precision, the $W$ boson mass with a 0.04\% 
precision. They together limit the mass of the SM Higgs to be smaller than 144 
GeV/$c^2$ at 95\% C.L..
CDF and D0 will continue to collect data (6-8 fb$^{-1}$ 
are 
expected by the end of Run 2) and to improve the precision on top and $W$ 
masses over the next few years.  

\section{Acknowledgments}

I would like to thank the SUSY07 organizers, for giving me the opportunity to 
give this presentation, and all of the CDF and D0 colleagues, who made 
possible to have these high quality physics results. In particular, my 
thanks to T. Bolton, G. Chiarelli, R. Erbacher, E. Halkiadakis, C. Hays, 
E. James, E. Shabalina, K. Tollefson, W. Wagner for their help and 
suggestions in preparing this talk and paper. All errors and omissions are 
mine.

\end{document}